\documentclass[%
 aps,prl,
 reprint,
 amsmath,amssymb,
 floatfix,superscriptaddress
]{revtex4-1}

\usepackage{graphicx}
\usepackage{dcolumn}
\usepackage{bm}
\usepackage{braket}
\usepackage{enumitem}

\usepackage{hyperref}
\hypersetup{
  colorlinks   = true, 
  urlcolor     = black, 
  linkcolor    = blue, 
  citecolor   = blue 
}

\setlist[itemize]{itemsep=0.1pt, topsep=6pt}

\usepackage{color}

\definecolor{cincinnati-red}{RGB}{190,0,0}

\begin{document}

\title{Emergent mode and bound states in single-component one-dimensional lattice fermionic systems}

\author{Yuchi He}
\thanks{These two authors contributed equally}
\affiliation{Department of Physics, Carnegie Mellon University, Pittsburgh, PA, 15213, USA}
\affiliation{Pittsburgh Quantum Institute, Pittsburgh, PA, 15260, USA}
\author{Binbin Tian}
\thanks{These two authors contributed equally}
\affiliation{Pittsburgh Quantum Institute, Pittsburgh, PA, 15260, USA}
\affiliation{Department of Physics and Astronomy, University of Pittsburgh, Pittsburgh, PA, 15260, USA}
\author{David Pekker}
\affiliation{Pittsburgh Quantum Institute, Pittsburgh, PA, 15260, USA}
\affiliation{Department of Physics and Astronomy, University of Pittsburgh, Pittsburgh, PA, 15260, USA}
\author{Roger S. K. Mong}
\affiliation{Pittsburgh Quantum Institute, Pittsburgh, PA, 15260, USA}
\affiliation{Department of Physics and Astronomy, University of Pittsburgh, Pittsburgh, PA, 15260, USA}

\begin{abstract}
We study the formation of bound states in a one-dimensional, single-component Fermi chain with attractive interactions. The phase diagram, computed from DMRG (density matrix renormalization group), shows not only a superfluid of paired fermions (pair phase) and a liquid of three-fermion bound states (trion phase), but also a phase with two gapless modes.  We show that the latter phase is described by a two-component Tomonaga-Luttinger liquid (TLL) theory, consisting of one charged and one neutral mode. We argue based on our numerical data, that the single, pair, and trion phases are descendants of the two-component TLL theory.  We speculate on the nature of the phase transitions amongst these phases.
\end{abstract}

\maketitle


Tomonaga-Luttinger liquid (TLL) theory captures the physics of many one-dimensional (1D) quantum systems such as spin chains, spin ladders, nanotubes~\cite{bockrath1999luttinger}, nanowires~\cite{deshpande2010electron}, and cold atoms confined to 1D tubes~\cite{kinoshita2004observation, capponi2008molecular,  gring2012relaxation, Yang2018, Salomon2018}.
In higher-dimensional systems, TLL is a tool that is often used, e.g.,\ in edge theory~\cite{wen} and coupled-wire constructions~\cite{KML, cazalilla2006interacting, Mross2016}. 

Recently, there has been significant interest in the study of 1D systems that cannot be described by the standard TLL theory~\cite{imambekov2009universal,imambekov2012one, mattioli2013cluster,dalmonte2015cluster, PhysRevLett.119.215301,SPV,JLSL}. In describing 1D interacting fermions, TLL theory naturally arises through bosonization that maps fermionic modes to bosonic modes. Nearby phases (i.e., descendants) such as charge density order appear as instabilities of the parent TLL theory~\cite{PhysRevLett.33.589, GS, affleck1988critical,penc1994phase,PhysRevLett.95.240402,PhysRevLett.103.215301,PhysRevLett.104.065301,roux2011multimer}.
This approach breaks down at the weak to strong pairing transition, i.e., the transition to the $p$-wave paired liquid. As recently pointed out in Ref.~\cite{KSH,RA}, the $p$-wave pairing phase cannot be described as a descendant phase of a single-mode TLL; instead the transition is described by an emergent mode theory, with the weak and strong pairing phases being descendants of this theory.
Which raises the question: what other phases, beyond $p$-wave pairing, can appear in one-component interacting fermions and how are these phases connected to some emergent-mode description?

In this paper, we investigate the formation of multifermion bound  states in 1D single-component systems. We perform DMRG numerics on a lattice model with finite-range interactions, and find liquids of singles, pairs, trions, etc.,\ in addition to  an extended phase with two gapless modes (2M phase). 
We unify these findings by constructing an effective theory with an emergent mode that characterizes the 2M phase, the descendants of this theory describe the liquid phases of single fermions as well as multifermion bound states (i.e., bound states of $2, 3, 4, \dots$ fermions).  Our construction [Eq.~\eqref{eq:general}] is not equivalent to the band-bending construction in Ref.~\cite{RA} (see supplement~\cite{supp})   but is similar to  Ref.~\cite{KSH} (see discussion at the end of the paper).

\textit{Microscopic model --}
We study the lattice Hamiltonian
\begin{align}
H=\sum_{i}\left[ -\frac{1}{2}\left(c^{\dagger}_{i}c_{i+1}+c^{\dagger}_{i+1}c_{i}\right)+ \sum_{m=1}^{3}V_{m}n_i n_{i+m}\right],
    \label{eq:latticeH}
\end{align}
where $c_i$ and $c^\dagger_i$ are the fermion annihilation and creation operators at lattice site $i$, $n_i=c^\dagger_i c_i$ is the number operator, and $V_m$ defines the shape of the fermion-fermion interaction potential.
We choose short-ranged attractive interactions ($V_1<0$ and $V_2<0$) to promote the formation of pairs and trions, but with $V_3>0$ to prevent phase separation~\cite{KSH}. 
To decrease the parameter space we restrict our attention to the subspace $V_1=V_2$. We expect that extending the range of attractive interactions will result in more liquid phases of multifermion bound states. For example, we demonstrate that extending the attractive interactions to three sites results in a quaternion liquid phase~\cite{supp}.

We use infinite-system DMRG (iDMRG)~\cite{White,mcculloch2008infinite,iDMRGpaper} to study the ground state properties of the Hamiltonian~\eqref{eq:latticeH} with a focus on the $1/5$ filling. 
The accuracy of iDMRG is controlled by the bond dimension $\chi$; the result becomes exact as $\chi\rightarrow\infty$~\cite{supp}. To identify the various phases, we use two types of diagnostics: central charge $c$ and various two-point correlators. 

We obtain $c$ as follows. We study the bipartite entanglement entropy $S$, i.e., the von Neumann entropy of DMRG ground state traced over either half of the system. Both $S$ and the correlation length $\xi$ are infinite for the true ground state, but are cut off by finite $\chi$. The manner in which these two variables diverge gives the central charge: $S = \frac{c}{6} \log(\xi)+\text{const}$~\cite{CC,pollmann2009theory}.

We also compute the single, pair, and trion two-point correlators
\begin{subequations}\label{eq:def_corr}\begin{align}
G_1(r)&=\big\langle c_i^\dag \, c_{i+r} \big\rangle, \\
G_2(r) &= \Braket{ (c_i c_{i+1})^\dag \, c_{i+r} c_{i+r+1} }, \\
G_3(r)&=\Braket{ (c_i c_{i+1} c_{i+2})^\dag \, c_{i+r} c_{i+r+1}c_{i+r+2}}.
\end{align}\end{subequations}
In the single phase, all correlators decay algebraically; in the pair phase, only $G_2$ decays algebraically while $G_1$ and $G_3$ decay exponentially; in the trion phase, $G_3$ is algebraic while $G_1$ and $G_2$ are exponential. This behavior implies that there is a bulk gap to adding a single fermion into the pair (trion) phase but no bulk gap to adding two (three) fermions.

\begin{figure}
	\includegraphics[width=8cm]{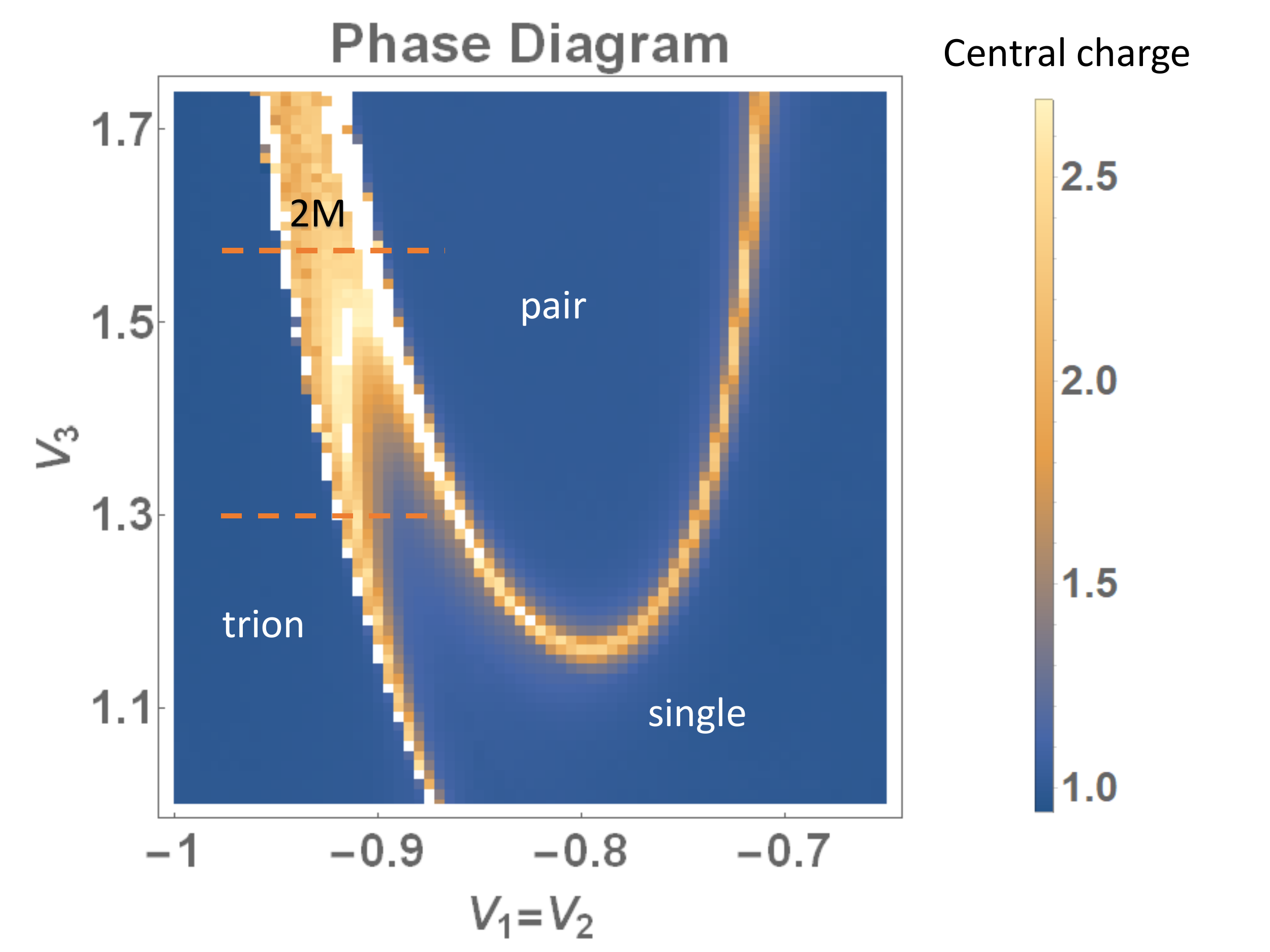}
    \caption{Central charge as a function of interactions in the lattice model \eqref{eq:latticeH} computed at filling fraction $1/5$. Four phases labeled are identified with further analysis of correlation functions. We have checked that all the reported phases exist for $V_{1}=2V_{2}$ and we expect the qualitative features of the phase diagram to hold for generic values of $V_1\neq V_2$ in the vicinity of $V_{1}=V_{2}$. 
    }
    \label{fig:pd}
\end{figure}

Figure~\ref{fig:pd} shows $c$ as a function of the interaction parameters $V_1=V_2$ and $V_3$. The blue regions denote the single-mode phases with $c=1$; we identify these as single, pair, and trion phases based on their two-point correlators [Eq.~\eqref{eq:def_corr}].
While we observe a direct transition between the pair and single phases~\cite{KSH,RA}, we do not find a direct transition between the pair and trion phases; instead we find an intermediate phase with $c\approx2$ which we call the 2M (two-mode) phase. The 2M phase neighbors all other phases and indicates a parent theory with an emergent mode, which enables a unified description of the multifermion bound-state phases and their transitions.

Our strategy for the remainder of the paper is as follows.  First, we introduce a field theory to describe the 2M phase, writing down its operators and free Hamiltonian.  Next, we introduce the possible interaction terms, and examine the descendant phases that result.  Finally, we show that these predictions are consistent with our numerics, justifying our theoretical model.

\textit{Theory of the emergent mode --\/}
Motivated by Refs.~\cite{KSH, RA} and our data, we introduce a theory with two modes. In this theory, the charge-$1$ operators in the lowest harmonic are~\footnote{We omit Klein factors for notational simplicity. }:
\begin{align}\begin{split}
	\psi_{0,\eta}^{(\pm)} &= e^{\pm i\theta_1}e^{i\theta_{0}+i\eta(\phi_{0}+k_{\text{F}}x)},
\\	\psi_{1,\eta}^{(\pm)} &= e^{ i\theta_{0}} e^{\pm i\theta_1 \pm i\eta(\phi_{1}+k^{\prime}x)},
	\label{eq:charge1}
\end{split}\end{align}
where $\eta = +1$ ($-1$) denotes a right (left) mover;
$\theta_\mu$ is the dual field of the compact bosonic field $\phi_\mu$ and satisfies $[\partial_x\theta_\mu(x), \phi_\nu(x')] = i\pi\delta_{\mu\nu}\delta(x-x')$. 
The charge is carried by the $\theta_0$ mode, while $\theta_1$ is neutral; as a result, $k_{\text{F}}$ is fixed by the density of electrons while $k'$ is a free parameter. 

The set of local physical operators can be generated via products of operators from Eq.~\eqref{eq:charge1}, i.e., $\big(\psi_{0,1}^+\big)^l \big(\psi_{0,-1}^+\big)^m \big(\psi_{1,1}^-\big)^n \cdots$.
[Note that the generators Eq.~\eqref{eq:charge1} are over-complete.]
As a result, primary operators of charge $q$ take the form:
\begin{align}\begin{split}
    c(x)^q
    &\sim \sum_{q_1,r_0,r_1} e^{i [q\theta_0 + q_1\theta_1 + r_0(\phi_0+k_{\text{F}} x) + r_1(\phi_1+k^\prime x)]} ,
\\  &\text{where $q_1 \equiv r_0 + r_1 \equiv q \pmod{2}$.}
    \label{eq:general}
\end{split}\end{align}
Due to the restrictions on the coefficients $q_1$, $r_0$, and $r_1$ of physical operators, we cannot simply treat this theory as a product of decoupled $\theta_0/\phi_0$ and $\theta_1/\phi_1$ theories.

\begin{table*}
\begin{center}
	\renewcommand{\arraystretch}{1.2}
  \begin{tabular}{ | c | c | c | c |c}
    \hline
    Locking term  & $\cos(2\theta_1)$ & $\cos(2k'x + 2\phi_1)$  & $\cos[(3k'-k_{\text{F}})x+3\phi_1-\phi_0]$ & $\dots$ \\ \hline
	Resulting phase & single & pair & trion& $\dots$ \\ \hline
    Single correlator $G_1(r)$  & $\sum_n \dfrac{\sin[(2n+1)k_{\text{F}}|r|]}{|r|^{(1/K+(2n+1)^2K)/2}}$ &  &  \\ \hline
    Pair correlator $G_2(r)$ & $\sum_n \dfrac{\cos[(2n)k_{\text{F}}|r|]}{|r|^{2/K+2n^2K}}$& $\sum_n \dfrac{\cos[(2n)\frac{k_{\text{F}}}{2}|r|]}{|r|^{(1/K+(2n)^2K)/2}}$ & \\ \hline
    Trion correlator $G_3(r)$ & $\sum_n \dfrac{\sin[(2n+1)k_{\text{F}}|r|]}{|r|^{(9/K+(2n+1)^2K)/2}}$& & $\sum_n  \dfrac{\sin[(2n+1)\frac{k_{\text{F}}}{3}|r|]}{|r|^{(1/K+(2n+1)^2K)/2}}$\\ \hline
  \end{tabular}
  \caption{Locking terms and correlators of single-mode phases. The first line lists interaction terms and the second line shows the corresponding phases when interaction terms get locked.  The remaining rows show the algebraic decay form of correlators $G_{1,2,3}$; the coefficient of each term is neglected for simplicity.
Figure~\ref{fig:exponent} shows the numeric data verifying the predicted dependence.
  }
  \label{table:correlators}
\end{center}
\end{table*}

The theory must obey charge conservation, and be invariant under both parity ($\phi_{0,1} \to -\phi_{0,1}$ and $x \to -x$) and time reversal ($\theta_{0,1} \to -\theta_{0,1}$, $i \to -i$, and $t \to -t$).
The kinetic part of the Hamiltonian takes the form:
\begin{align}
	\mathcal{H}_{\text{KE}} = \sum_{\mu,\nu} \big[ A_{\mu\nu}(\partial_x\theta_\mu)(\partial_x\theta_\nu) + B_{\mu\nu}(\partial_x\phi_\mu)(\partial_x\phi_\nu) \big].
    \label{eq:2MKinetic}
\end{align}
$\mathcal{H}_\text{KE}$ describes a two-mode TLL, which we later demonstrate to be consistent with the 2M phase found in the numerics.

\textit{Single-mode phases as descendants of the 2M theory --}

The single-mode phases (single, pair, trion, ...) are constructed by introducing locking terms, shown in Table~\ref{table:correlators}, to the Hamiltonian~\eqref{eq:2MKinetic}. 
For a term to appear, it must be of the form of Eq.~\eqref{eq:general} with $q = 0$, and also respect parity and time reversal. At large interaction strength, some of these terms may ``lock"~\cite{roux2011multimer}, taking an expectation value and reducing the theory to a one-component TLL.

Our analysis for the locking terms follows~\cite{gogolin2004bosonization}. For an interaction term to lock, it should have no oscillation (i.e., $x$ dependence), which places constraints on the Fermi momenta.
For each locking term of the form $\cos \Lambda$, we find linear combinations of the $\theta$'s and $\phi$'s that commute with $\Lambda$.  Among this set we find a conjugate pair which we denote as $\theta_+$ and $\phi_+$.
The set of gapless operators is then generated by $e^{i \Lambda}$, $e^{i\theta_+}$, and $e^{i\phi_+}$, and must be a subset of Eq.~\eqref{eq:general}~\footnote{Although the operators $e^{ib\Lambda}$ are not gapless, they are long-ranged correlated (ordered).}.
We show that the minimal (unit) charges for these operators are indeed $q_\text{min} = 1$, 2, 3 for the single, pair, trion phases respectively, from the given locking terms.
We extend our analysis to arbitrary $q_\text{min}$ in the supplement~\cite{supp}. 

We first analyze the locking term $\cos(2\theta_1)$ which induces the single phase.
The gapless mode is described by the dual fields $\theta_+ = \theta_0$ and $\phi_+ = \phi_0$.
Thus the gapless operators take the form $c(x) \sim \sum e^{ia\theta_1}e^{i\theta_0}e^{i(2n+1)(\phi_0+k_{\text{F}}x)}$ where $n$ is an integer and $a$ an odd integer.
(The dual field $\phi_1$ is disordered and cannot appear here.)
As $e^{i\theta_1}$ is a constant, $c(x)$ reduces to the standard bosonization form of a fermion mode~\cite{haldane1981effective,roux2011multimer}.

Next, we show that the locking term $\cos(2\phi_1+2k'x)$ induces the pair phase.
Notably, for this term to lock we must enforce $k' = 0$.
As $\theta_1$ is disordered, it cannot appear in a gapless operator, i.e., $q_1 = 0$.
From the parity relation \eqref{eq:general}, we see that $q$ must be an even integer and thus the single and trion correlators decay exponentially.
Letting $\theta_+=2\theta_0$ and $\phi_+=\phi_0/2$, we recover the standard bosonization expansion of a boson mode~\cite{haldane1981effective,roux2011multimer} for the pair operator: $c(x)^{2} \sim b(x) \approx \sum e^{i\theta_+} e^{i(2n)(\phi_{+}+k_Bx)}$ with $n\in\mathbb{Z}$ and $k_B = k_{\text{F}}/2$.
We interpret this descendant theory as a TLL of fermion pairs, with the density of pairs being half of the density of elementary fermions.

Finally, we address the locking term $\Lambda = 3(\phi_1+k^\prime x)-(\phi_0+k_{\text{F}} x)$ which yields the trion phase while fixing $k' = k_{\text{F}}/3$.
As $\Lambda$ commutes with $\theta_+ = 3\theta_0 + \theta_1$ and $\phi_+ = \phi_1$, the gapless operators take the form $c(x)^q \sim \sum e^{i(q/3)\theta_+} e^{ia(\phi_+ + k'x) + ibL}$.
Mapping the expression to Eq.~\eqref{eq:general},
we get $q_1 = q/3$, $r_0 = -b$, and $r_1 = a + 3b$; we determine the consistency conditions $q/3, a, b \in \mathbb{Z}$ and $a \equiv q \pmod{2}$.
Hence for any gapless operator, $q$ must be a multiple of 3, which implies exponential decay of $G_1$ and $G_2$.
The trion operator expansion reduces to:
$c(x)^3 \sim \sum e^{i\theta_+} e^{i(2n+1)(\phi_+ + k'x)}$, where $k' = k_{\text{F}}/3$ is the Fermi wave vector of the trions and $n$ is an integer.

Within the low energy theory for each of the three single TLL mode phases, $c(x)^{q_\text{min}}$ admits a standard bosonization expansion in terms of $\theta_+$ and $\phi_+$.
The effective Hamiltonian is thus
\begin{align}\label{plussectoreft}
	\mathcal{H}_{+} &= \frac{v_{+}}{2\pi}\big[K(\partial_{x}\theta_{+})^2 + \frac{1}{K}(\partial_{x}\phi_{+})^2\big] ,
\end{align}
where $K$ is the Luttinger parameter.

\begin{figure}%
\flushleft
    \includegraphics[width=9cm]{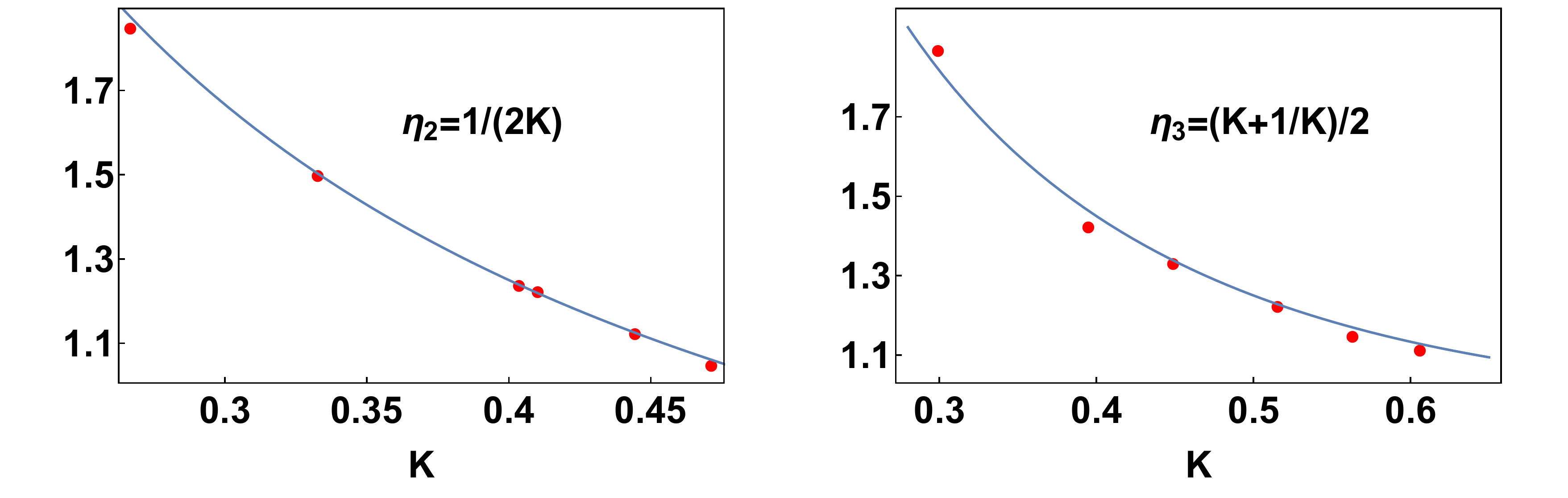}
    \caption{Verification of predicted  decay exponents.
    Left: leading decay exponent ($\eta_2$) of pair correlator in pair phase at $V_1=V_2=-0.8$, $V_3=1.4$.
    Right: leading decay exponent ($\eta_3$)  of trion correlator in trion phase at $V_1=V_2=-1$, $V_3=1.4$.
    The two lines are predictions from TLL theory (cf.\ Tab.~\ref{table:correlators}),
    $\eta_2 = \frac{1}{2K}$ and $\eta_3 = \frac{1}{2}(K+\frac{1}{K})$.
    The values of Luttinger parameter $K$ are extracted from the neutral sector \cite{supp}.  In order to cover a larger range of $K$, we use DMRG data from fillings (left to right) $\frac{1}{5}$, $\frac{1}{6}$, ..., $\frac{1}{10}$.
    }
    \label{fig:exponent}
\end{figure}

\textit{Fourier spectra of the correlators --\/}
The long-distance behavior of the correlation functions of gapless operators can be written as a sum of algebraically decaying terms of the form
\begin{align}
	\frac{\cos(k_\text{osc}|r|+\varphi)}{|r|^\eta} .
    \label{eq:decayform}
\end{align}
Our theory puts a restriction on the allowed values of $k_\text{osc}$ in the 2M phase and the single-mode phases.
Table~\ref{table:correlators} summarizes the long distance behavior of the correlation functions within the single-mode phases;
observe that the (leading) decay exponents $\eta$ of all harmonics $k_\textrm{osc}$ depend only on the Luttinger parameter $K$; this is verified in Fig.~\ref{fig:exponent}.

\begin{figure}%
\includegraphics[width=\columnwidth]{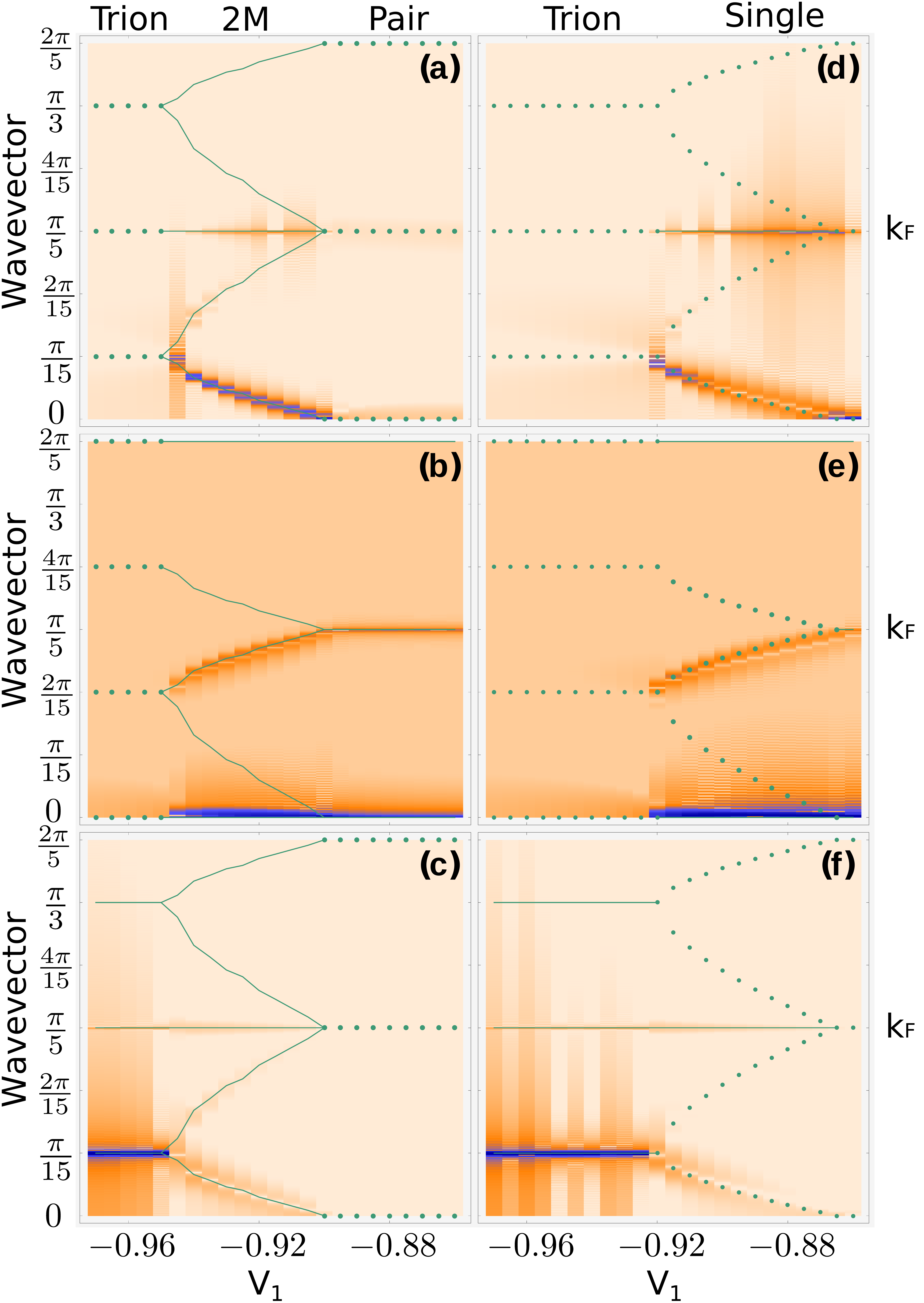}
    \caption{%
    Spectra $G_1$, $G_2$, and $G_3$ (from top to bottom) as a function of wave vector and interaction strength ($V_1=V_2$), showing agreement of peak locations between DMRG and theory.
	The data is taken at cuts shown in Fig.~\ref{fig:pd}.
    Plots (a)–(c) taken at $V_3=1.56$ show the trion, 2M, and pair phases; plots (d)–(f) taken at $V_3=1.3$ show the trion and single phases (with a possible 2M phase in between).
   Darker (blue) colors represent larger values of amplitudes. The peak in the data of $G_1$, which continuously varies between 0 and $k_{\text{F}}/3$ in the 2M/single phase, is identified as $k^{\prime}$. The lines added to the color plot are theoretic predictions with the determined parameter $k^{\prime}$.  The solid lines denote several long-distance $k_\textrm{osc}$ associated with algebraic-decay; the dotted lines denote several exponential-decay ``peaks", which are possibly visible if the decay length scale is large.
   The parameters for the plots are explained in the supplement~\cite{supp}.}
    \label{fig:cuts}%
\end{figure}

To connect the effective theory to our microscopic model, we compare the $k_\textrm{osc}$ in correlation functions obtained from field theory and DMRG.
We perform Fourier transforms on the correlation functions $G_{1,2,3}(r)$ and take the $n$\textsuperscript{th} derivative, such that terms of the form Eq.~\eqref{eq:decayform} with $\eta<n+1$ will show a divergent peak at $k_\textrm{osc}$.
We then match the set of predicted oscillation wave vectors to peaks in the Fourier transforms.
Figure~\ref{fig:cuts} presents the correlation functions along cuts at constant $V_3$.
Panels (a)–(c) show a cut through the trion, 2M and pair phases; while panels (d)–(f) cut from trion to single phase (with a possible intervening 2M phase).

In the 2M phase, both modes are gapless and the allowed $k_\text{osc}$'s are given by the oscillatory part of $c(x)^q$ in Eq.~\eqref{eq:general}:
\begin{align}
	k_\text{osc} = r_0 k_{\text{F}} + r_1 k'.
    \label{eq:unicorn}
\end{align}
For $G_1$ and $G_3$, $r_0+r_1$ is odd, hence the first several $k_\text{osc}$ are $k'$, $k_{\text{F}}$, $k_{\text{F}}\pm2k'$, and $2k_{\text{F}}\pm k'$.  For $G_2$, $r_0+r_1$ is even and so $k_\text{osc} = 0$, $k_{\text{F}}\pm k'$, $2k'$, $2k_{\text{F}}$, etc.
These wave vectors are fitted to numerical data and are marked by the dotted lines in Fig.~\ref{fig:cuts}.
As $k_{\text{F}}$ is fixed, $k'$ is the only fitting parameter at each point of phase space.
In the 2M region of panels (a)–(c), we observe unambiguous peaks at the predicted wave vectors.
(In the numerics, we have not resolved peaks at some of the predicted $k_\text{osc}$, as these peaks are too weak or have large exponent $\eta$.)

A key feature of the DMRG data in the 2M phase is that $k'$ varies continuously between the two limiting values: $k'=k_{\text{F}}/3$ on the trion side and $k'=0$ on the pair side. The variation of this wave vector is a clear sign of a neutral emergent mode and confirms our effective two-mode TLL~\footnote{We discuss the incompatibility of this result and Ref.~\cite{RA} in the supplement~\cite{supp}.}.

In the single-mode phases, some of the peaks found for the 2M phase persist while others are no longer divergent as modes become gapped out.
The trion phase is characterized by the absence of singular behavior in panels (a), (b), (d), (e) as $G_{1,2}$ decay exponentially.
We observe that $G_3$ decays algebraically with peaks in Figs.~\ref{fig:cuts}(c) and (f) at odd multiples of $k' = k_{\text{F}}/3$.
In the pair phase, $k'=0$ and only $G_2$ shows divergent peaks at multiples of $k_{\text{F}}$, as predicted.  [The features at $0,k_{\text{F}}$ in panel (a) are not divergent and broadened out due to $G_1$ being gapped. Deep in the pair phase, they become invisible.]
Finally, for the single phase all three correlators are algebraically decaying [Figs.~\ref{fig:cuts}(d)–(f)] with peaks at multiples of $k_{\text{F}}$.
Remarkably, we also observe exponentially decaying features at the moving $k'$ which are remnants of the 2M parent theory.

\textit{Phase transitions --}
There are five potential phase transitions in our phase diagram.
The locking mechanisms give hints about the possible phase transitions, which we discuss in relation to our data.
\begin{itemize}
\item Single-pair transition.
The transition is controlled by the competition between the terms $\cos(2\theta_1)$ and $\cos(2\phi_1)$, and results in a quantum Ising transition~\cite{LGN,KSH,RA,sitte2009emergent,alberton2017fate}. In the supplement~\cite{supp}, we provide definitive evidence that the single-pair transition is Ising via finite-$\chi$ scaling.
\item 2M-single transition.
This transition is driven by the term $\cos(2\theta_1)$, and is likely a Berezinskii-Kosterlitz-Thouless transition. 
\item 2M-pair/trion transition.
Both 2M-to-pair and 2M-to-trion transitions are accompanied by $k'$ reaching a commensurate value. This suggests a commensurate-incommensurate transition.
\item Single-trion transition.
We are unable to determine if there is a direct transition between the trion and single phase, or whether there is an intervening 2M phase which extends down as $V_3$ is decreased.  In both cases, our numerical analysis suggests a (at least one) first-order transition (see the supplement~\cite{supp}). 
\end{itemize}

\textit{Discussion --} In summary we find conclusive evidence for an emergent mode in a one-dimensional attractive fermion chain. This emergent mode results in the formation of a stable 2M phase with two Fermi surfaces. We argue that the multifermion bound-state liquids are not descendants of the single-mode TLL phase but are rather descendants of this 2M phase. Here the 2M parent theory is written as a mixture of charged/neutral modes.  Curiously, we can also rewrite the theory in terms of a mixture of charge-1/charge-$2$ modes~\cite{KSH}, or more generally charge-$n$/charge-$(n+1)$ modes.

The two ingredients required to realize the proposed phenomenology are (1) confining the fermions to one dimension and (2) controlling the form of the interaction potential between the fermions. In the setting of solid-state systems the two ingredients could be realized in nanowires made of superconducting semiconductors~\cite{mourik2012signatures,chen2017j,deng2016majorana,Annadi2018,cheng2015electron,cheng2016tunable}.
%
%
%
%
In ultracold atoms, confinement could be provided by either optical lattices~\cite{kinoshita2004observation,Yang2018,bloch2008bloch}, or atom chips~\cite{gring2012relaxation} and tunable long-range interaction by the use of dipolar interactions~\cite{baranov2012condensed,gross2017quantum},
or Rydberg-state-mediated interactions~\cite{zeiher2017coherent}.

The 1D systems studied here can also be used to construct higher dimensional topological phases via the coupled-wire construction~\cite{KSH,TK, KML,mong2014universal}.
TLL enriched by emergent mode(s) may give a pathway to a wide range of new phases in condensed matter.

\begin{acknowledgments}
We acknowledge enlightening discussions with J.~Levy.  This work was supported by the Charles E.~Kaufman foundation KA2017-91784 and NSF PIRE-1743717.
\end{acknowledgments}


\clearpage
\appendix

\section{Numerical method for DMRG calculation}

We use the standard two-site-update iDMRG algorithm~\cite{White,mcculloch2008infinite,iDMRGpaper} to find the ground states of the Hamiltonian~\eqref{eq:latticeH} at various filling ratios.
For the calculation (1/5 filling) presented in the main text, we use a unit cell with 30 fermion sites.  The $\mathrm{U}(1)$ charge is explicitly conserved in the DMRG simulations.

In the calculations, we use various bond dimensions $\chi$ for the purpose of checking convergence and doing finite $\chi$ scaling. The phase diagram Fig.~\ref{fig:pd} is computed using $\chi = \{80, 160, 300\}$. The Fourier spectra Fig.~\ref{fig:cuts} is plotted using the $\chi=300$ data.

The correlation lengths $\xi(\chi)$ are used to compute the central charge $c$ and also determine whether correlators are algebraically or exponentially decaying.
For each charge sector $Q$, we can compute $\xi_Q(\chi)$--the length scale for correlators of form $\braket{A^\dag(0) B(r)}$ where $A,B$ are charge-$Q$ operators.
As $\chi$ increases, $\xi_Q(\chi)$ goes to infinity as long as that charge-$Q$ sector is gapless.  If instead, the charge-$Q$ sector is gapped, then $\xi_Q(\chi)$ saturates to its physical value $\xi_Q(\infty)$.
We find for TLL of $q_\text{min}$-bound states, the charge $Q$-sectors are gapless only if $Q$ is a multiple of $q_\text{min}$.
We use this method to determine $q_\text{min}$, i.e., whether the phase is single, pair, or trion. In practice, if $\xi_{Q=jq_\text{min}}(\chi), j=0,1,2...$ reach some values significantly larger than other charge sectors whose $\xi_{Q}(\chi)$ tend to saturate, we claim the formation of liquid of $q_\text{min}$-bound states.  Our data with $\chi=300$ can practically confirm a gapped sector with up to about a hundred correlation length.
The neutral sector ($Q=0$) is always gapless in our phase diagram and the corresponding $\xi_0(\chi)$ is used to extract $c$~\cite{pollmann2009theory}~\footnote{The formula $S = \frac{c}{6} \log(\xi_0)+\text{const}$ in the main text, was derived in the context of a CFT, where there is a single velocity. For two-component phases, we still use this formula as an operational definition of the central charge despite the likely possibility of mismatched Fermi velocities.}.

Translational invariance is not preserved exactly by iDMRG at finite $\chi$. Instead, the magnitude of violation decays algebraically with increasing $\chi$. When we compute correlators, we average over sites $r_1$ in one unit cell:  $G(r)= \frac{1}{\text{unit cell size}}\sum_{r_1} G(r_1,r_1+r)$. We also do this when calculating entanglement entropy. The averaging improves the quality of the data.

\subsection{Numerical methods for Fourier spectra (Figure~\ref{fig:cuts})}
As shown in the main text, the Fourier transform of correlators reveals the leading $k_\text{osc}$ as peaks in the spectra. The Fourier spectra Fig.~\ref{fig:cuts} are plotted using data from ground states with parameters along two cuts. The spectra is calculated as the ``$n$\textsuperscript{th}" derivative of Fourier transform on the correlation functions $G_{1,2,3}(r)$:
\begin{align}
\tilde{G}_{1,2,3}^{n}(q)=\frac{1}{\sqrt{2r_\text{trunc}}}\sum_{r=-r_\text{trunc}}^{r_\text{trunc}} e^{iqr} |r|^n G_{1,2,3}(r),
\end{align}

where the value of $r_{\text{trunc}}$ is set to be several times of the correlation length $\xi_0(\chi)$.

Any harmonic component (Eq.~\eqref{eq:decayform}) of correlators  with algebraic decay exponent $\eta<n+1$ will in principle show a divergent peak at the corresponding $k_\textrm{osc}$. 
To make the subleading $k_\text{osc}$ visible, we choose $n=2$, $2.5$, $2.8$ for $G_1$, $G_2$ and $G_3$.

\section{DMRG evidence for a quaternion phase}

The Hamiltonian Eq.~\eqref{eq:latticeH} can be extended to have longer ranges of attractive interaction, which is expected to give rise to more $q_{\text{min}}$-bound state. 
We use the same numeric method to study the range-four model with $V_{1}=V_{2}=V_{3}<0$, $V_{4}>0$ in 1/5 filling to find a quaternion phase.

\begin{figure}%
\includegraphics[width=8.6cm]{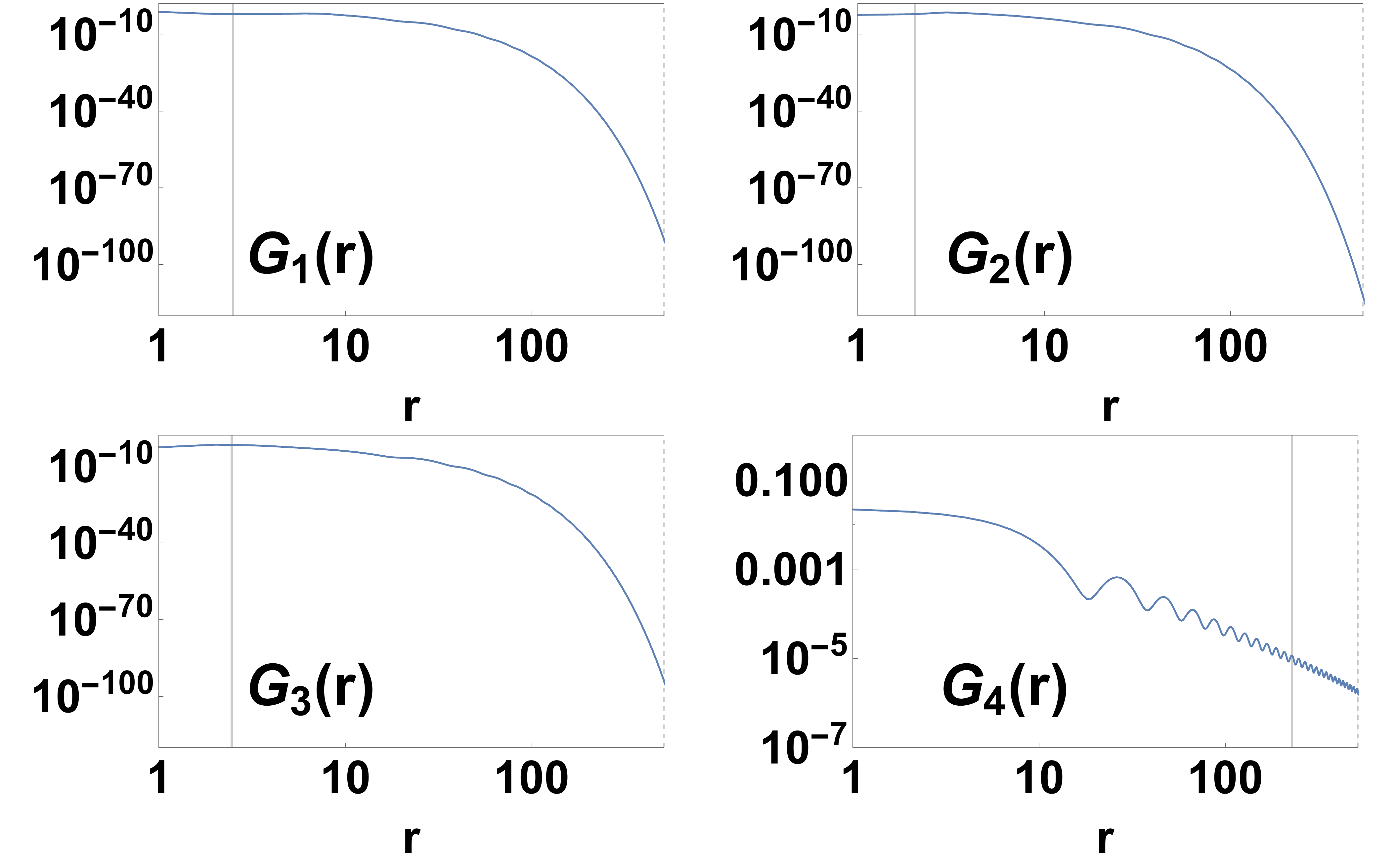}
\caption{%
	Correlators in quaternion phase. The parameters of Hamiltonian are $V_{1}=V_{2}=V_{3}=-0.7$, $V_
    {4}=1.7$. The DMRG bound dimension $\chi=600$.
    The solid vertical lines in sub-figures of $G_1(r)$, $G_2(r)$, $G_3(r)$, and $G_4(r)$ are the DMRG correlation length $\xi_q(\chi)$ in charge $q=1$, 2, 3, and 4 sectors respectively.
    For $q=1$, 2 and 3, $\xi_q(\chi)$ saturates to a small value $\approx 2$. In contrast, $\xi_4(\chi)$ increases with increasing $\chi$, with $\xi_4(\chi=600) \approx 227$. 
The data indicates that $G_1(r)$, $G_2(r)$ and $G_3(r)$ decay exponentially while $G_4(r)$ decays algebraically. }%
    \label{fig:quaternioncorrelators}%
\end{figure}

To show the minimal charge ($q_{\text{min}}$) for gapless charged sector is 4, we demonstrate the correlators $G_{1,2,3,4}$ of a state in the quaternion phases in Fig.~\ref{fig:quaternioncorrelators}, where $G_{4}$ is the quaternion correlator:
\begin{align}
G_4(r)=\Braket{ (c_i c_{i+1} c_{i+2}c_{i+3})^\dag \, c_{i+r} c_{i+r+1}c_{i+r+2}c_{i+r+3}}.
\end{align}

\begin{figure}%
\includegraphics[width=6cm]{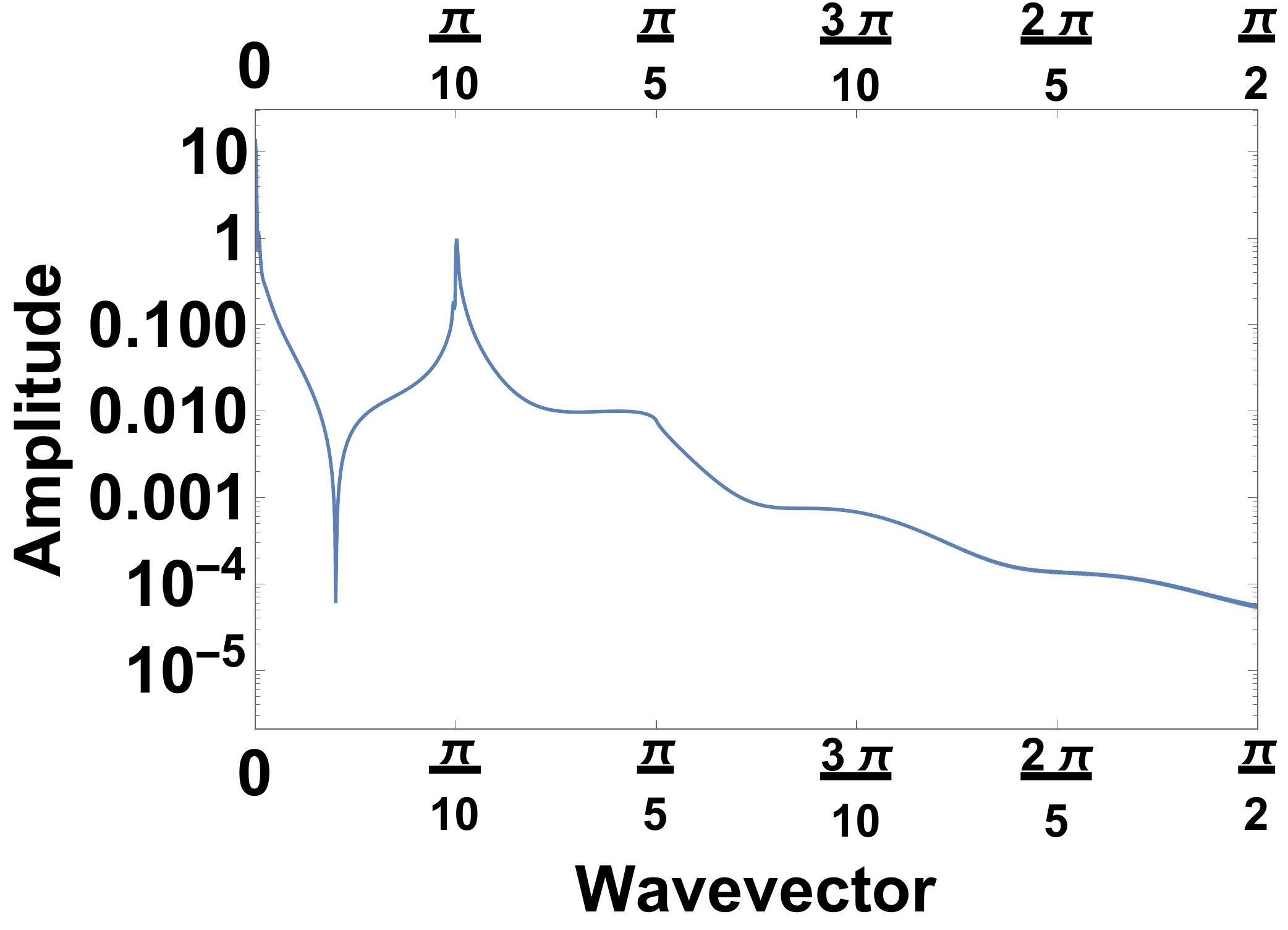}
\caption{Fourier spectra of quaternion correlator ($G_{4}$ of Fig.~\ref{fig:quaternioncorrelators}) in a quaternion phase. The oscillatory wavevectors ($k_{\text{osc}}$) are located at even multiples of $\pi/20$. With a choice ``derivative'' $n=2$, the first three $k_{\text{osc}}$ can be seen in this plot as peaks or step.}
    \label{fig:quaternionFourier}
\end{figure}

In Fig.~\ref{fig:quaternionFourier}, we show the Fourier spectrum of $G_{4}$ with the choice of derivative $n=2$ (see the last section). As the filling is 1/5, the quaternion density is 1/20. As quaternions are bosons, the spectrum shows peaks or steps at the even multiples (0, 2, 4) of $\pi/20$.

\section{Numeric method for self consistency check of the decay exponents in correlators for pair and trion phases Fig.~\ref{fig:exponent}}

In this section, we demonstrate that the pair and trion phases are well-described by a single-mode TLL theory.
We do this by extracting the decay exponents of the correlators from our data, and comparing them to prediction from TLL theory.  In particular, we first extract the Luttinger parameter $K$ from the scaling dimension of the  leading oscillatory term in the density operator.
We then compute the leading decay exponent ($\eta_{2,3}$) of $G_{2,3}$, and show that they follow the relation $\eta(K)$ as predicted by TLL theory.

We first show that $K$ can be read out from charge-density wave quasi-order.
The density-density correlator is defined as: 
\begin{align}
G_0(r)=\braket{n_in_{i+r}}-\braket{n_i}^2,
\end{align}
where $n_i=c^\dag_i c_{i}$.
TLL theory predicts $n(x) \sim \braket{n}+\partial_x\phi_{+}/\pi+\sum_{m\neq 0} e^{i2m(\phi_{+}+k_F x/q_\text{min})} + \dots$ and thus the long-distance behavior of $G_0(r)$ reads:
\begin{align}
G_{0}(r) = \frac{1}{r^2} + \sum_{m\neq 0}  \frac{\cos(2mk_{\text{F}}r/q_\text{min})}{|r|^{2m^2K}} + \dots,
\end{align}
where the coefficient of each term is neglected. The term with $m=1$ is the leading oscillatory (quasi-CDW order) term with decay exponent $2K$.  Thus the leading scaling dimension of the quasi-CDW order is $K$.

\begin{figure}
    \centering
    \includegraphics[width=10cm]{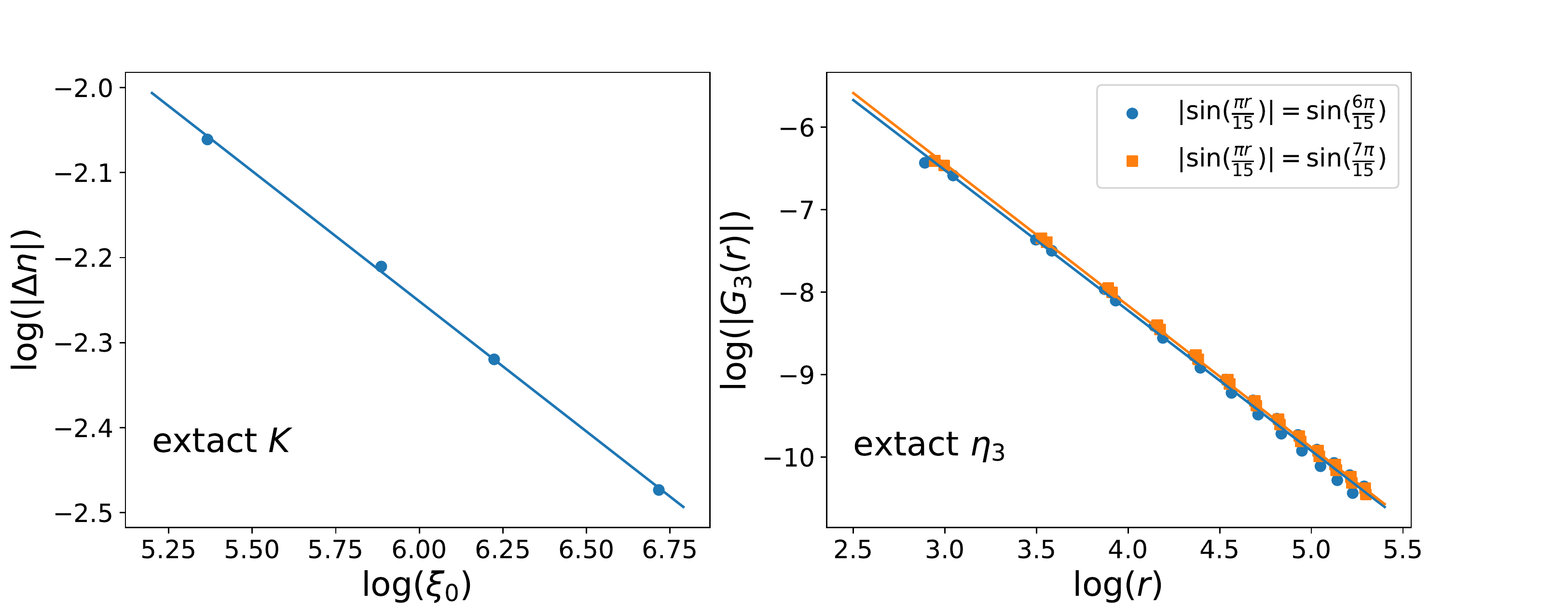}
    \caption{Methods for extracting exponents plotted in Fig.~\ref{fig:exponent}. This example shows the extraction for the left most point of the right figure.}
    \label{extarct}
\end{figure}

While it is possible to extract $K$ by computing $G_0$ from our data, we choose to fit $K$  by taking advantage of the artificial long-range charge-density-wave order induced by iDMRG.
The method~\footnote{The method works for any quasi-orders which are not forced to vanish by the MPS ansatz} is as follows.  When the unit cell of iDMRG is commensurate to the oscillatory vector(s) of the quasi-charge-density-wave order, there is a corresponding artificial long-range density-wave order induced from finite bond dimension $\chi$, the amplitude of which decays with $\chi$ and the correlation length $\xi_0(\chi)$.
Specifically, its amplitude scales as
\begin{align}
	\Braket{n\Big(\frac{2k_F}{q_{\text{min}}}\Big)}_{\chi} \propto \xi_0(\chi)^{-K},
\end{align}
where $\Braket{n(2k_F/q_{\text{min}})} = \sum_j e^{2ik_F/q_{\text{min}}j} \Braket{n_j}$.
For our plots, we use the peak-to-peak amplitude ($|\Delta n|$ in Fig.~\ref{extarct}) of the density profile as a substitute for $\braket{n(2k_\text{F}/q_{\text{min}})}$.

TLL theory [cf.\ Tab.~\ref{table:correlators}] gives the prediction of leading term of $G_{2,3}$ of pair and trion phases respectively as: 
\begin{align}
\begin{split}
	G_2(r) &= \frac{1}{|r|^{\eta_2}}+\dots,
\\	G_3(r) &= \frac{\sin(k_{\text{F}}|r|/3)}{|r|^{\eta_3}}+\dots,
\end{split}    
\end{align}
with
\begin{align}\label{eta23}
\begin{split}
	\eta_2 &= \frac{1}{2K},
\\	\eta_3 &= \frac{1}{2}\left(K+\frac{1}{K}\right) .
\end{split}    
\end{align}
We numerically fit $\eta_{2,3}$ directly from the data of $G_{2,3}$.  Here, we introduce the fitting procedures. $G_{2}(r)$ has an non-oscillatory leading term. For $G_{3}(r)$, its leading term has a oscillatory factor. To use log-log plot to fit the decay exponent $\eta_3$, we can divide $|G_{3}(r)|$  into 8 branches corresponding to the 8 possible values of the factors $|\sin(\frac{r\pi}{15})|$.  Note that one branch is flat as $|\sin(\frac{0\pi}{15})|=0$. We extract $\eta_3$ using two branches with the largest factors. We illustrate the fitting procedure in Fig.~\ref{extarct}. Note that we ignore the subleading term in the fitting, whose exponent is $9K+1/K$ (Table I), which is close to $\eta_3$ when $K$ is small. This should be partly responsible for the relatively large fitting errors of $k_F=\pi/15$ data and its slight deviation from the theoretical curve.

Figs.~\ref{fig:exponent}(a) and (b) show $\eta_{2,3}$ vs.\ $K$ for the pair and trion phases respectively.
We see that the data points (red) agree with Eqs.~\eqref{eta23} (blue line), which shows that the liquids are well described by single-mode TLL theory.
We also note that the extracted Luttinger parameters $K$ are all smaller than 1; this indicates the effective interaction between pairs or trions is repulsive.

\section{Why our numerical data is inconsistent with the band bending theory of \texorpdfstring{Ref.~\onlinecite{RA}}{Ruhman, Altman}}

\begin{figure}
\includegraphics[width=0.5\textwidth]{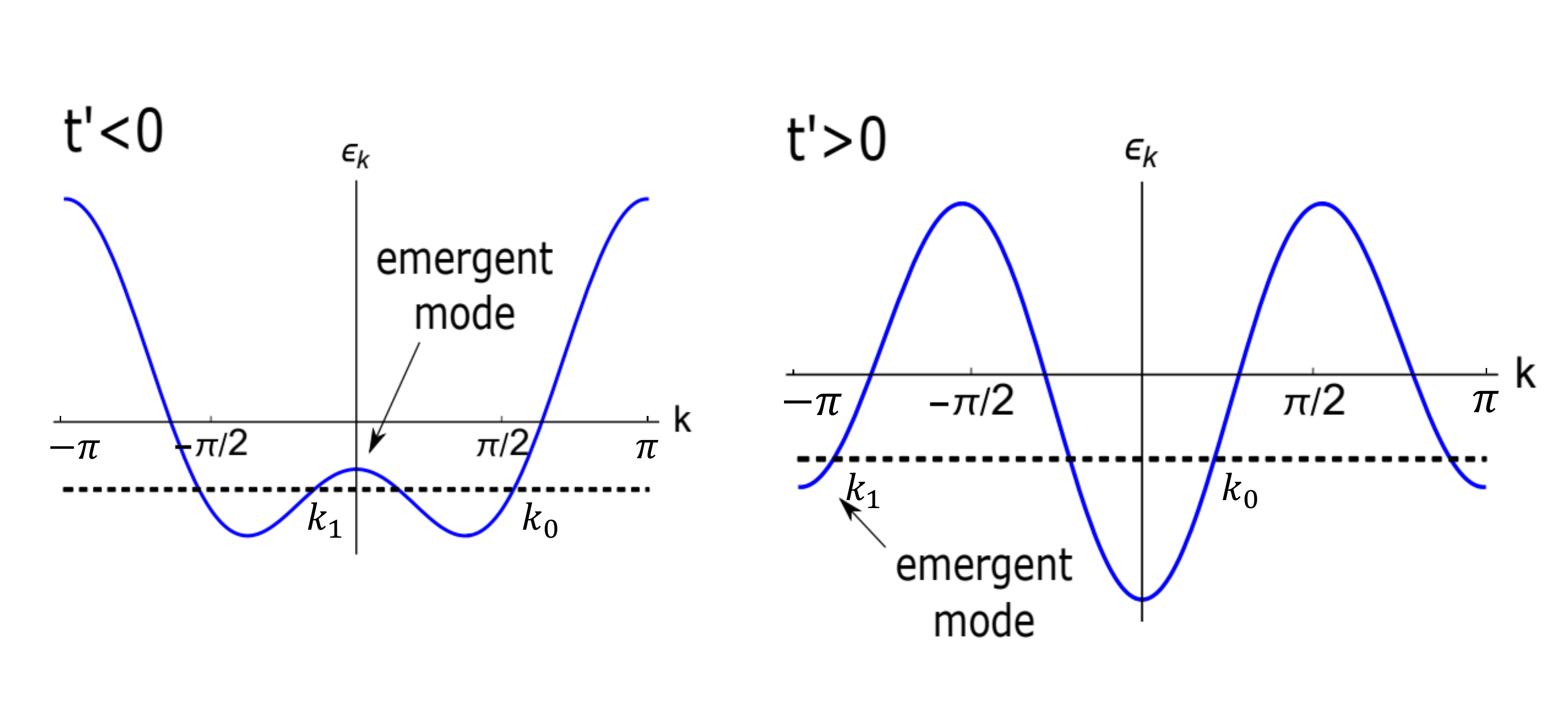}
\caption{Band bending pictures from Ref.~\onlinecite{RA} where $k_0$ and $k_1$ are the right moving wavevectors. To satisfy Luttinger's theorem we need $k_0+k_1=k_\text{F}$ (left) and $k_0+k_1+\pi=k_\text{F}$ (right).
}
\label{fig:RAbandbending}
\end{figure}

In the effective theory of Ref.~\onlinecite{RA} two distinct Fermi-vectors appear due to the bending of the band (as shown in Fig.~\ref{fig:RAbandbending}). The two Fermi-vectors $k_0$ and $k_1$ should satisfy Luttinger's theorem: $k_0+k_1=k_F$ or $k_0+k_1+\pi=k_F$. In our numerical calculations we focus on filling fraction of $1/5$ and $k_F=\pi/5$. However, our numerical data shows that in the 2M phase one of the Fermi-vectors moves from $k_1=0$ to $k_1=k_F/3$ while the other Fermi-vector stays constant at $k_0=\pi/5$. The band bending picture would suggest that both Fermi points move together, either  $k_0+k_1=\pi/5$ or $k_0+k_1+\pi=\pi/5$. Hence, our numerical results are inconsistent with the predictions of the effective theory of Ref.~\onlinecite{RA}. It should be noted that Ref.~\onlinecite{RA} deals with a different microscopic model, which may have a different effective field theory.

\section{A theory of bound states composed of an arbitrary number of fermions}
In this section, we show that all liquid phases of $q_{\text{min}}$-bound states can be formed by the locking of appropriate vertex terms in the 2M parent theory. We derive the low energy expansion of the $c(x)^{q_{\text{min}}}$ operator (corresponding to the $q_{\text{min}}$-bound state) and show that it corresponds to the standard TLL theory of a charge $q_{\text{min}}$ particle.

We first reproduce Eq.~\eqref{eq:general} as the starting point, which gives the form of a charge-$q$ physical operator in the 2M theory:
\begin{align}\begin{split}
c(x)^q
&\sim \sum_{q_1,r_0,r_1} e^{i \left(q\theta_0 + q_1\theta_1 + r_0(\phi_0+k_{\text{F}} x) + r_1(\phi_1+k^\prime x)\right)} ,
\\ &\text{where $q_1 \equiv r_0 + r_1 \equiv q \pmod{2}$.}
\label{eq:appendixgeneral}
\end{split}\end{align}

The central goal is to show the locking of some vertex operators reduces 2M expansion (Eq.~\eqref{eq:appendixgeneral}) to liquid phases of $q_{\text{min}}$- bound states. For the sake of constraining the form of vertex operators relevant to this goal, it is convenient to notice a field-redefinition invaraince of Eq.~\eqref{eq:appendixgeneral}.

Consider the following redefinition of the fields, which preserves $\theta_0$, $\phi_0$ as a charge one field, and $\theta_1$, $\phi_1$ as a neutral field:
\begin{align}\label{eq:field transform}
\begin{split}
	\phi_1 &\to \pm\phi_1-j\phi_0 ,
\\	\theta_0 &\to  \theta_0+j\theta_1, \\
\phi_0 &\to \phi_0,\\
\theta_1 &\to \pm \theta_1,\\
k^{\prime} &\to \pm k^{\prime}-jk_{\text{F}}, \\
k_F &\to k_F.
\end{split}
\end{align}
For even $j$, the redefinition Eq.~\eqref{eq:field transform} leaves Eq.~\eqref{eq:appendixgeneral} invariant. (For odd $n$, we get a different but equivalent form of the operator expansion for the 2M phase.)
Observe that it is always possible to transform the fields such that $k'$ is non-negative.

We then decide the form of vertex terms relevant to this goal.

Recall from the main text: all these TLL of $q$-bound states are described by a single-mode TLL theory.
To reduce the number of gapless mode of the 2M theory down to one, there should be a single locking term, i.e., $\propto \cos(\Lambda)$, where $\Lambda$ is a linear combination of fields and dual fields.
The term needs to respect the following symmetries~\footnote{As the interaction terms in the boson theory come from a four-fermion term, the Klein factors are absent and neither the transform of Klein factor under P,T introduces a minus sign for those terms}:
\begin{itemize}
\item Parity: $\phi_{0,1} \to -\phi_{0,1}$ and $x \to -x$.
\item Time-reversal: $\theta_{0,1} \to -\theta_{0,1}$ and $t \to -t$.
\item Charge $\mathrm{U}(1)$ symmetry: $\theta_0 \to \theta_0 + \varphi$; i.e., the term cannot include the field $\theta_0$.
\end{itemize}

With the above constrain, the allowed term takes the form $\cos(2p\theta_1)$ or $\cos\big( q\phi_1-p\phi_0+(qk^{\prime}-pk_{\text{F}})x \big)$ with $p \equiv q \pmod{2}$. For the second kinds of terms, by the invariance via fields redefinition Eq.~\eqref{eq:field transform} with even $j$, we can constrain $0<p\leq q$ and $k'\geq0 $ without lose of generality.
Note that we do not consider Umklapp scattering caused by the possible presence of lattice.

Some allowed locking linear combinations in the cosine may differ by an integer multiple, for example  $\cos(2\theta_1)$ vs.\ $\cos(4\theta_1)$. Among those terms, the term with the smallest coefficient is the most relevant and we further constrain our discussion on the locking of such terms.
We have shown that  $\cos(2\theta_1)$ leads to the single-TLL phase, we focus on the remaining terms $\cos\Lambda$ for constructing TLL of $q_{\text{min}}$-bound states.
We summarize these terms as follows:
\begin{subequations}\label{eq:locking}
\begin{align}
&\cos(2\theta_1),
	\label{eq:locking_single}\\[1ex]
\begin{split}
&\cos\big[ (2n+1)\phi_1-(2m+1)\phi_0 \\&\qquad+((2n+1)k^{\prime}-(2m+1)k_{\text{F}})x \big],\end{split}
	\label{eq:locking1}\\[1ex]
&\cos\big[ 2n\phi_1-2m\phi_0+(2nk^{\prime}-2m k_{\text{F}})x \big], 
	\label{eq:locking2}
\end{align}
\end{subequations}
where  in Eq~\eqref{eq:locking1}  $2m+1$ is coprime to $2n+1$; in Eq.~\eqref{eq:locking2}, $m$ is coprime to $n$, and $m+n\equiv 1\pmod{2}$.
Here $n \geq m$ are non-negative integers.

In the following, we construct the TLL of $q_{\text{min}}$-bound states from the 2M expansion Eq.~\eqref{eq:appendixgeneral} with the vertex terms Eq.~\eqref{eq:locking}.

We have already demonstrated in the main text that the locking term Eq.~\eqref{eq:locking_single} results in the single phase; $q_\text{min}=1$.

The locking of Eq.~\eqref{eq:locking1} always reduces the 2M theory to a TLL of $(2n+1)$-bound states.
This also places a constraint on $k^{\prime}$: namely $(2n+1)k^{\prime}-(2m+1)k_{\text{F}}=0$. We follow the procedure described in the main text to find the gapless degree of freedom and denote it by $+$. By the commutation condition $[\theta_+,\Lambda]=0$, we find a $\theta_+=(2n+1)\theta_0+(2m+1)\theta_1$, which carries $2n+1$ unit charge. Consequently, the gapless operators must carry integer multiples of charge $2n+1$.
To prove that the low energy bosonization expansion of $c^{2n+1}$ is takes the form of the standard fermion expansion, it's sufficient to prove some allowed linear combination of $\phi_0$ and $\phi_1$ gives $\phi_+$, which is dual to $\theta_+$. To be specific, we want to find $\phi_+=a\phi_0-b\phi_1$ such that $[\partial_x\theta_+(x,t),\phi_+(x^{\prime},t)]=i\pi\delta(x-x^{\prime})$.
(The constraint on $a$, $b$ is $a+b\equiv 1\pmod{2}$, which is derived from  Eq.\eqref{eq:appendixgeneral}.)
The existence of the solution $a$ and $b$ is guaranteed by: for two coprime odd numbers $2n+1$ and $2m+1$, there are always two coprime numbers $a$ and $b$ such that $a(2n+1)-b(2m+1)=1$, where $a+b\equiv 1\pmod{2}$. Note that $ k_{\text{F}}x/(2n+1)$ is attached to this $\phi_+$ and the coefficients of $\phi_+$ are all odd numbers. Then we obtain the standard bosonization of charge-(2n+1) fermion: $c(x)^{2n+1} \sim \sum e^{i\theta_+} e^{i(2j+1)(\phi_+ + k'x)}$, where $k' = k_{\text{F}}/(2n+1)$ is the Fermi wavevector of the $(2n+1)$-bound states TLL and $j$ is an integer.

For the locking term Eq.~\eqref{eq:locking2}, the descendant theory is a TLL of $(2n)$-bound states.
Similar to the above analysis for $(2n+1)$-bound states, we find $\theta_+=(2n)\theta_0+(2m)\theta_1$, which carries charge $2n$. $2n$ is minimal charge for gapless operators because $n\theta_0+m\theta_1$ is not allowed as the $\theta$ sector of any physical operator.
[By construction, $m+n\equiv 1\pmod{2}$, while the constraint for physical operators is $m+n\equiv 0 \pmod{2}$.]
It's then sufficient to prove some allowed linear combination [odd-odd or even-even given by Eq.~\eqref{eq:appendixgeneral}] of $\phi_0$ and $\phi_1$ gives $2\phi_+$.
Indeed, this is true because for any two coprime numbers $n$ and $m$ obeying $m+n\equiv 1\pmod{2}$, there is always two coprime odd numbers $a$ and $b$ such that $a n-b m=1$. Note that $k_{\text{F}}x/(2n)$ is attached to  $\phi_+$ and the coefficients of $\phi_+$ are all even numbers. Then we obtain the standard bosonization of charge-(2n) boson: $c(x)^{2n} \sim \sum e^{i\theta_+} e^{i(2j)(\phi_{+}+k_Bx)}$ with $j\in\mathbb{Z}$ and $k_B = k_{\text{F}}/(2n)$.

\section{DMRG data for single-pair phase transition}
\begin{figure}%
\centering
    {\includegraphics[width=6cm]{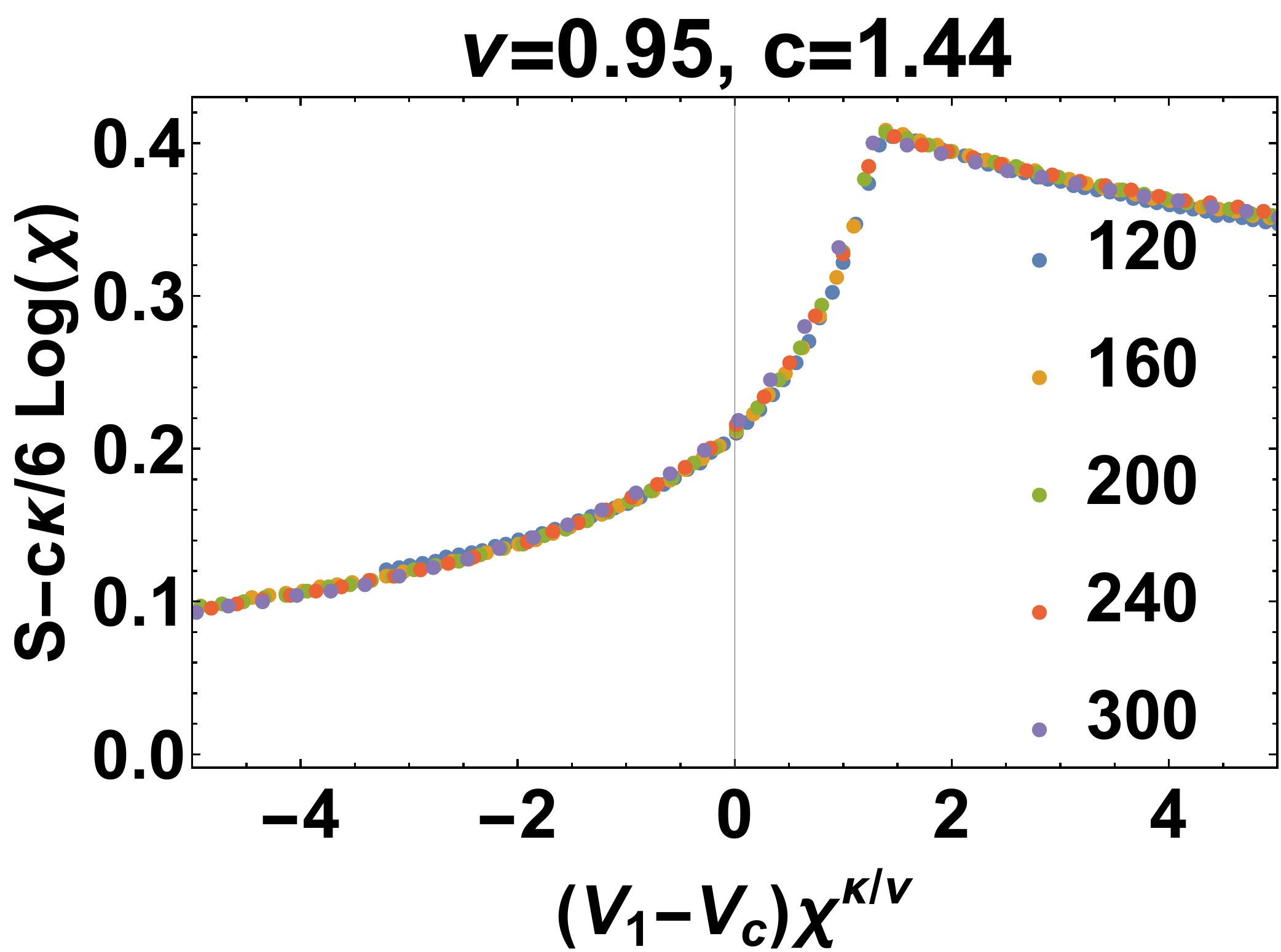} }%
    \caption{Single-pair transition: collapse of the entanglement entropy ($S$) data computed as a function of the tuning parameter ($V_1=V_2$) for various bond dimensions ($\chi$ -- indicated by different color points as labeled) using the scaling ansatz Eq.~\eqref{twocftansatz}. The collapse is used to extract the central charge $c$ and the correlation-length critical exponent $\nu$ at the phase transition. The data is taken from the cut at $V_3=1.25$. }%
    \label{fig:transitions}%
\end{figure}

\begin{figure}%
\centering
    {\includegraphics[width=6cm]{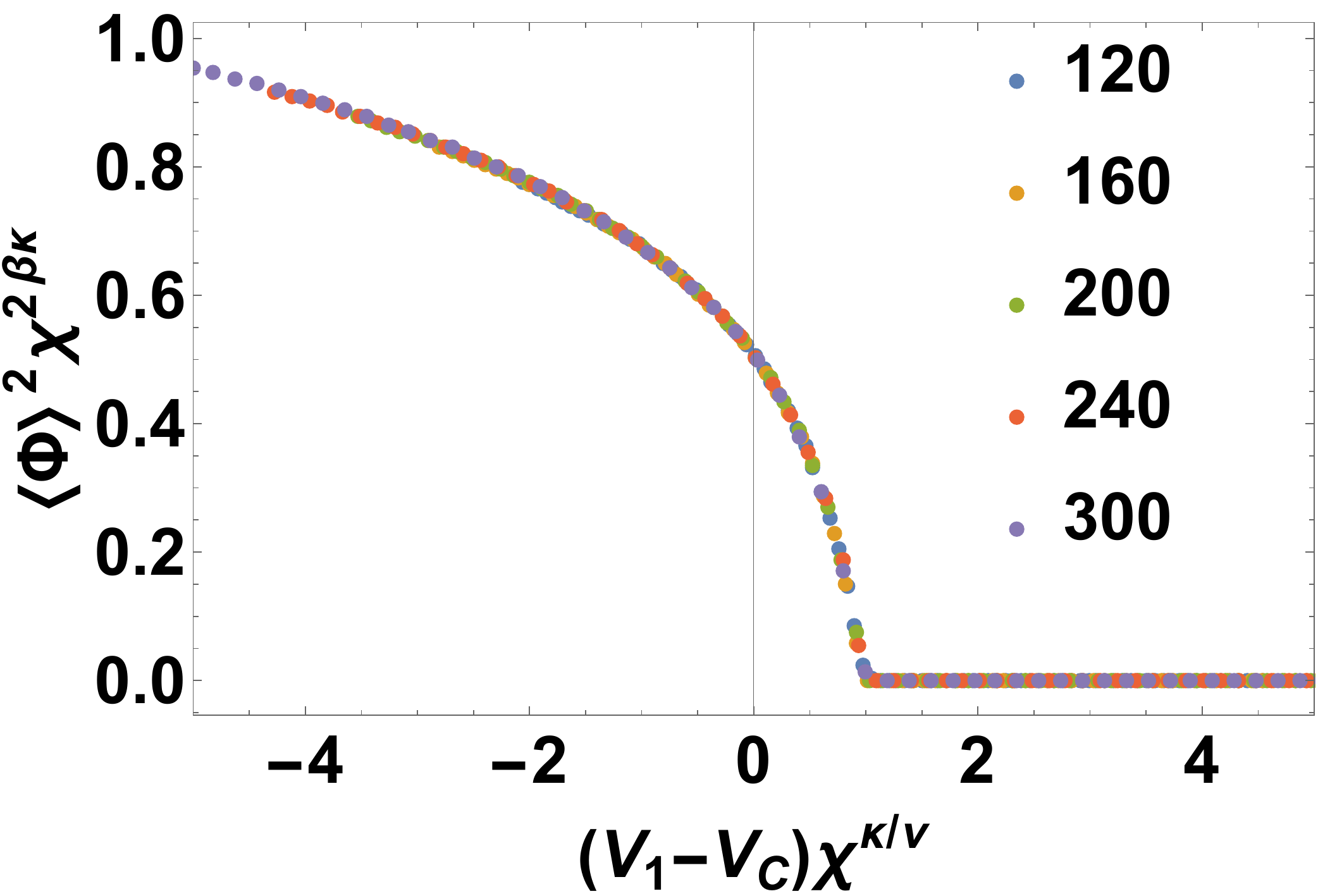} }%
    \caption{Single-pair transition: collapse of the disorder parameter ($\Phi$) computed as a function of the tuning parameter ($V_1=V_2$) for various  bond dimensions ($\chi$ -- indicated by different color points as labeled)  using the scaling ansatz Eq.~\eqref{jw}. The cut is the same as Fig.~\ref{fig:transitions}. The value of $\braket{\Phi}^2$ is extrapolated from $\braket{\Phi(i+r)\Phi(i)}$ with large $r$ as well as unit cell averaging over $i$, period averaging over $r$ with period 10.  Unlike Fig.~\ref{fig:transitions},  we use the exact Ising critical exponents for the collapse and find the collapse is good. }%
    \label{fig:transitionsJW}%
 \end{figure}

In this section we investigate the single-pair phase transition. We establish that this is an Ising transition by (1) measuring the central charge at the critical point, (2) measuring the correlation length critical exponent $\nu$, and (3) identifying the order parameter for the pair phase and finding its scaling dimension $\beta$.  Our technical approach is to perform a two-parameter scaling collapse on our DMRG data, where the parameters are detuning $V_1-V_C$ and bond dimension $\chi$.

We begin by extracting the central charge at the critical point and the correlation length critical exponent by analyzing the entanglement entropy in the vicinity of the critical point. Using the finite $\chi$ scaling of CFT states~\cite{pollmann2009theory}, we collapse the entanglement entropy data from various $\chi$ near the pair-single transition. The collapse ansatz is as follows:
\begin{align}\label{twocftansatz}
S-\frac{c\kappa}{6}\log(\chi)=f\left(\left(V_1-V_c\right)\chi^{\kappa/\nu}\right),
\end{align}
where $\kappa=\frac{6}{c(1+\sqrt{\frac{12}{c}})}$ and $\nu$ is the correlation-length critical exponent. We note that $\nu=1$ and $c=1/2$ for the Ising transition. The best collapse of our DMRG data yields $\nu=0.95$ and $c=1.44$. These values differ by less than 5\% from the Ising critical point values, once we realize that in addition to the CFT that corresponds to the Ising transition, there is a background free boson CFT with central charge 1, and thus  $c=1+1/2=3/2$. We suspect that the 5\% deviation is due to the subleading corrections to the entanglement entropy scaling, which have been ignored in the collapse anstaz.

We also analyze the expectation value of the vortex operator $\Phi(x)$ and perform the corresponding collapse of the finite $\chi$ data (Fig.~\ref{fig:transitionsJW}). The vortex operator $\Phi(x)$, is neutral and incurs a $\pi$ phase upon braiding (in spacetime) with fermions; i.e., it anticommutes with fermion operators on its left and commutes with fermion operators on its right. Expanding in terms of primary fields, $\Phi(x)$ has the following operator expansion:
\begin{align}\begin{split}
\Phi(x)
&\sim \sum_{r_0,r_1} e^{i \left(r_0(\phi_0+k_{\text{F}} x) + r_1(\phi_1+k^\prime x)\right)} ,
\\ &\text{where $r_0 + r_1$ is odd.}
\label{eq:vortexgeneral}
\end{split}\end{align}
The operator $\cos(\phi_1+k^{\prime}x)$, which is one of the lowest harmonic terms in $\Phi(x)$, acquires long-range order in the pair phase, because $\cos(2\phi_1)$ is locked and $k^{\prime}=0$. In the context of the Ising transition, $\cos(\phi_1)$ corresponds to the disordered parameter, which gains an an expectation value on the disordered side of the transition with the magnetization critical exponent $\beta$.

In our lattice model, the vortex operator corresponds to: $\Phi(x)=\prod_{j<x}(-1)^{n_j}$.  Its two-point correlator is well defined and measurable in iDMRG.
Its large distance limit gives: $\lim_{r\rightarrow \infty} \braket{\Phi(i+r)\Phi(i)}=\braket{\Phi}^2$.  We measure $\braket{\Phi}^2$ near the phase transition point for various bond dimensions $\chi$ and the collapse the data using the following ansatz:
\begin{align}\label{jw}
\Braket{\Phi}^2\chi^{2\beta\kappa}=g\left(\left(V_1-V_c\right)\chi^{\kappa/\nu}\right).
\end{align}
In Fig.~\ref{fig:transitionsJW}, we plug in the exact Ising exponents $\beta=\frac{1}{8}$, $\nu=1$, and $\kappa=\frac{6}{c(1+\sqrt{\frac{12}{c}})}$ with $c=3/2$ to obtain good collapse of the iDMRG data near the Ising transition.  

\section{DMRG data at the interface of single and trion phases}
\begin{figure}%
\centering
    {\includegraphics[width=8cm]{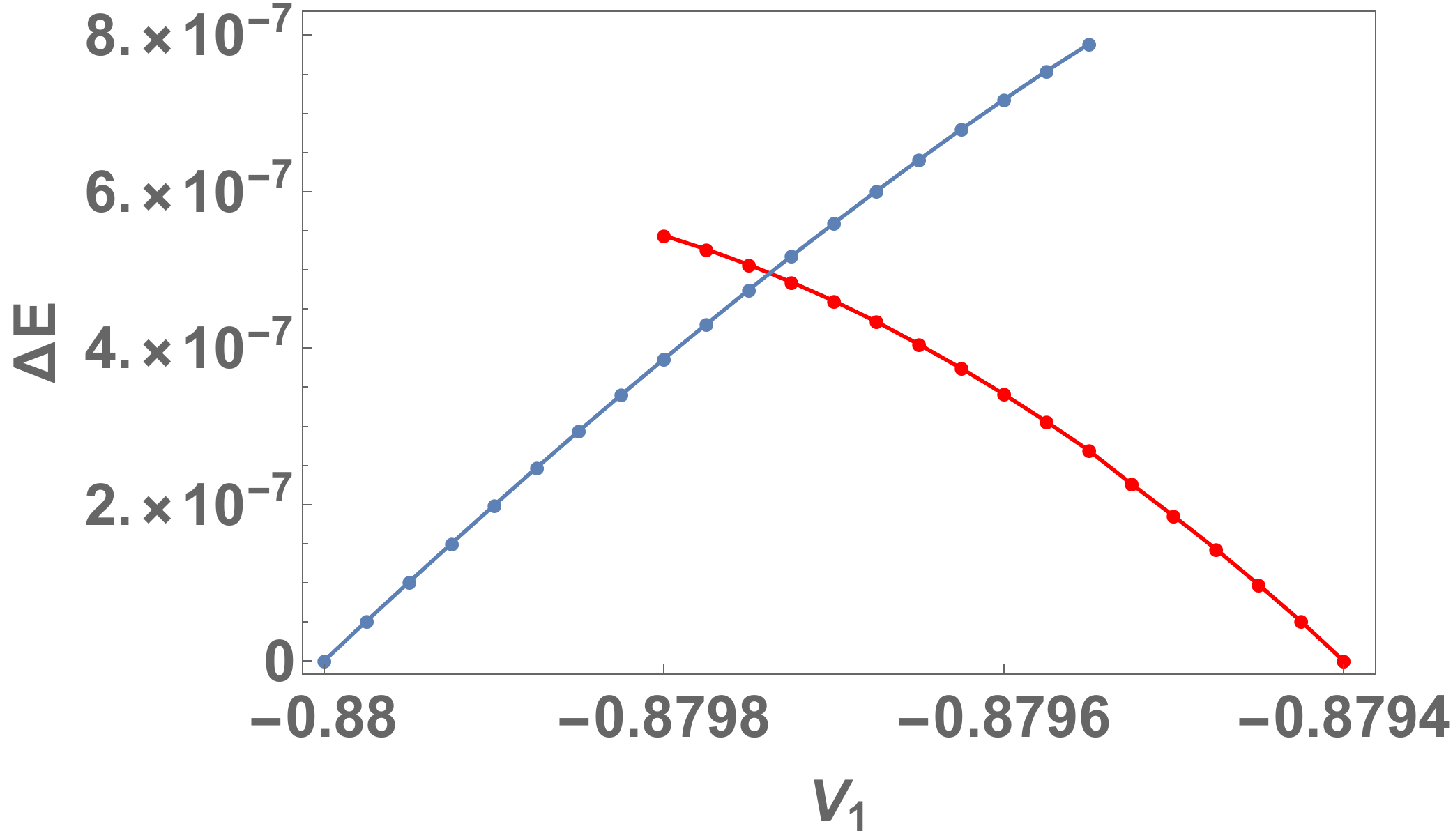} }%
    \caption{DMRG ground state energy (after subtracting a linear function, see text) along a cut at the interface of single (right) and trion (left) phases. The bond dimension is $\chi=300$. 
These two sets of data points correspond to the energy of two phases on either of the transition. 
There is a region where the two curves have overlap in the parameter ($V_1$) space. In this region, the data point with lower energy is the ground state, while the data point with higher energy indicates a metastable states.
This metastability is an artifact of DMRG which occurs near a first-order transition.}%
    \label{fig:singletrionE}
\end{figure}

We also analyze the iDMRG data along the cut $V_3=1.02$ which goes from the trion to the single phase. In Fig.~\ref{fig:singletrionE} we plot $\Delta E(V_1)=E(V_1) - (a V_1+b)$, the ground state energy minus a linear component, as a function of the tuning parameter $V_1$. We subtract the linear component to make the kink in the ground state energy easier to visualize. We see that $\Delta E$ has different slopes on trion/single side, which is a signature of first-order transition. However, we cannot rule out the possibility that there is an intermediate 2M phase between the trion and the single phases.

\bibliography{mybib}

\end{document}